# Liquid Reconfigurable Stealth Window Constructed by Metamaterial Absorber

Xiangkun Kong[1,2], Weihao Lin[1], Xuemeng Wang[1], Lei Xing[1], Shunliu Jiang[1], Lingqi Kong[1]

[1] Key Laboratory of Radar Imaging and Microwave Photonics, Ministry of Education, Nanjing University of Aeronautics and Astronautics, Nanjing 210016, People's Republic of China
2. State Key Laboratory of Millimeter Waves of Southeast University, Nanjing 210096, China

E-mail: xkkong@nuaa.edu.cn
E-mail: weihaolin@nuaa.edu.cn

**Abstract**

In this paper, a liquid reconfigurable stealth window constructed by metamaterial absorber at microwave band is proposed. The stealth window consists of an anti-reflection glass with indium tin oxide (ITO) as resistive film and a liquid container made of polymethyl methacrylate (PMMA). Since the materials constituting the window are all transparent, the metamaterials that can be switched through the liquid control system can always maintain high light transmission. The proposal can obtain a transmission passband from 2.3 GHz to 5 GHz with low insertion loss, especially at 2.45 GHz and 5 GHz with the insertion loss of the passband reach 0.51 *dB* and 0.99 *dB*, by alcohol drainage. It can also reflect electromagnetic waves at 2.45 GHz and absorb them from 4.5 GHz to 10.5 GHz with a strong absorptivity over 90% by alcohol injection, exhibiting the reconfigurable electromagnetic characteristic of switching between transmission state and absorption state. Furthermore, the proposed absorber shows its good transmission/absorption performance under different polarizations and obtains absorptivity over 90% when alcohol injection in an oblique incidence of 50°. Finally, the prototype window has been fabricated to demonstrate the validity of the proposed structure, which indicates that the proposal presents significant implications for smart stealth systems and WLAN communication that require switching of working states in a complex electromagnetic environment.

## 1. Introduction

Metamaterials defined as materials which are not existing in nature are artificial materials providing invisible characteristics in nature, such as negative refraction, perfect absorption, and perfect transmission, through the periodical arrangement of meta-atoms made by the size smaller than incident electromagnetic by about 1/3-1/5 [1]-[4]. As people pay more and more attention to the practical value of metamaterial, metamaterial absorbers have gradually been researched because of their absorption performance. One kind of classical absorber, the Salisbury screen, plays an important role to reduce the Radar Cross Section (RCS) in the microwave region [5]-[8]. However, due to the limitation of quarter wavelength resonance condition, the narrow bandwidth of the Salisbury screen limits its practical application under the certain case, hence many designs have been proposed to broaden the bandwidth of the absorbers, such as Jaumann screen absorbers [8]-[9] using multilayered configurations, circuit analogous absorbers [10]-[12] integrating resistive frequency selective surface and high-impedance surface based on Salisbury screen [13]-[14]. To meet the needs of the realistic situations, a miniaturized design about absorbers with broad bandwidth is proposed [15], making full use of structural space. For realizing the purpose of the true stealth system of the radome to reduce the RCS [16]-[17], metamaterial absorbers are usually applied to design the new type radome, obtaining the transmission at the passband but also absorb the out-of-band waves to reduce the bistatic RCS. Such a low RCS radome based on FSS and metamaterial absorbers is named as frequency selective rasorber (FSR) [18]-[20]. And in [21], the meander-line square loop and lumped resistivities were used by Yu and Luo to create a dual-polarized band-absorptive FSR with miniaturization, having the electromagnetic wave absorbed in the certain bandwidth and making a transparent window below the absorption band.

With the change of application requirement and the production technology progress, metamaterial absorbers with new characteristics, such as miniaturization and transparency, are designed, making them apply to more places. For example, the transparent metamaterial absorbers can be applied as a window in secure buildings. In recent years, Indium-tin-oxide (ITO) with commercial availability, low cost and good mechanical flexibility is applied as a resistive film in transparent metamaterial absorbers [22]. And the resistive film made from ITO can be designed to different shapes such as square shape and circular shape, but also can be attached to different dielectric materials such as anti-reflection glass and Polyethylene Terephthalate(PET) according to practical requirements. Novel absorbers with optical transparency were designed to reduce the RCS in a broader bandwidth [23]-[25]. Moreover, water has been put more and more attention in the design of water antennas [26] and metamaterials because of its unique characteristics such as transparency, a high value of permittivity and dielectric loss [27]. Water is applied as a substrate or primary resonant elements to realize metamaterial absorbers with visible transparent characteristics and broader absorption bandwidth [28]-[30].

The metamaterial structures mentioned above are all passive because their characteristics or functions are constant once they are designed or fabricated. Therefore, active metamaterial absorber or FSR is researched and designed in recent years because of their flexibilities, making that they can be applied in a more complex electromagnetic environment. Through the electrical control of varactors and PIN diodes, the active frequency selective surface (AFSS) can show the switching of transmission, reflection or absorption at the single/multiple frequencies of interest [31-34]. Unlike electrical control, by using the liquid's fluidity and the dispersion characteristics of the permittivity, our research group proposed a water-based reconfigurable FSR with a thermally tunable absorption band [27].

In this paper, a liquid transparent reconfigurable stealth window is proposed, which is the most unique. An anti-reflection glass with ITO as resistive film, a liquid container made from PMMA and 75% alcohol constructs this new design. The visible transparent characteristic of the materials mentioned above and the application of only one ITO layer make this proposal optically transparent. Moreover, it can switch from a bandpass response for 2.45G and 5G WLAN application to perfect wideband absorption, realized by alcohol drainage or injection. The realization of reconfigurable function takes advantage of the liquid's fluidity and the dispersion characteristics of the permittivity. Also, it is polarization-insensitive due to the symmetry of structure and steady for large incident angles. Finally, the prototype has been fabricated and measured to demonstrate the validity of the proposed structure.

## 2.Window structure and simulation results

The liquid reconfigurable stealth window is a sandwich structure consisting of an anti-reflection glass layer, an air layer, two layers of the liquid container and an alcohol layer, as shown in Figure 1. The top layer is an anti-reflection glass and the ITO conductive film structure of which the square resistance value is 25 $\Omega/Sq$ is printed on it. As shown in Figure 2(a), the radius of the circular ITO conductive film is $r$. The thicknesses of the anti-reflection glass, air substrate, alcohol substrate and monolayer of the liquid container, shown in Figure 2(b), are $d_g$, $d_a$, $d_w$ and $d_p$, respectively. The geometrical parameters are $p = 28\,mm$, $r = 6.5\,mm$, $d_g = 0.7\,mm$, $d_a = 1\,mm$, $d_w = 7.5\,mm$, $d_p = 3\,mm$.

Full-wave electromagnetic simulation is used to simulate the unit structure of the window. In the numerical simulation, the permittivity of the anti-reflection glass and the PMMA are 5.5 and 2.65×(1- $j$ 0.012), respectively. The permittivity of alcohol at electromagnetic waves can be described by the Debye formula, which is as a function of frequency $f$, temperature $T$ and substance concentration $S$ as follows[35]:

$$\varepsilon(f,T,S) = \varepsilon_\infty(T,S) + \frac{\beta_0(T,S)}{\alpha_0(T,S) + j2\pi f} \quad (1)$$

Where the parameters $\varepsilon_\infty$, $\alpha_0$ and $\beta_0$ are dependent on temperature $T$ and substance concentration $S$

$$\varepsilon_\infty(T,S) = \sum_{m+n=0}^{3} C_{\varepsilon mn} S^m T \quad (2)$$

$$\alpha_0(T,S) = \sum_{m+n=0}^{3} C_{\alpha mn} S^m T^n \quad (3)$$

$$\beta_0(T,S) = \sum_{m+n=0}^{3} C_{\beta mn} S^m T^n \quad (4)$$

**Table 1**. Parameters of the fit (2), (3), (4) for alcohol.

| m, n | $C_{\varepsilon mn}$ | m, n | $C_{\alpha mn}$ | m, n | $C_{\beta mn}$ |
|------|------|------|------|------|------|
| 0, 0 | 8.840 | 0, 0 | 57.752 | 0, 0 | 4.238e+03 |
| 0, 1 | -0.082 | 0, 1 | 2.191 | 0, 1 | 1.867e+02 |
| 0, 2 | -0.004 | 0, 2 | 0.015 | 0, 2 | 0.787 |
| 0, 3 | 2.765e-05 | 0, 3 | 2.548e-04 | 0, 3 | 0.012 |
| 1, 0 | 0.085 | 1, 0 | -2.534 | 1, 0 | -2.052e+02 |
| 1, 1 | 0.006 | 1, 1 | -0.052 | 1, 1 | -5.876 |
| 1, 2 | 1.179e-05 | 1, 2 | -3.515e-04 | 1, 2 | -0.024 |
| 2, 0 | -0.002 | 2, 0 | 0.049 | 2, 0 | 4.133 |
| 2, 1 | -4.805e-05 | 2, 1 | 3.821e-04 | 2, 1 | 0.050 |
| 3, 0 | 9.464e-06 | 3, 0 | -3.154e-04 | 3, 0 | -0.028 |

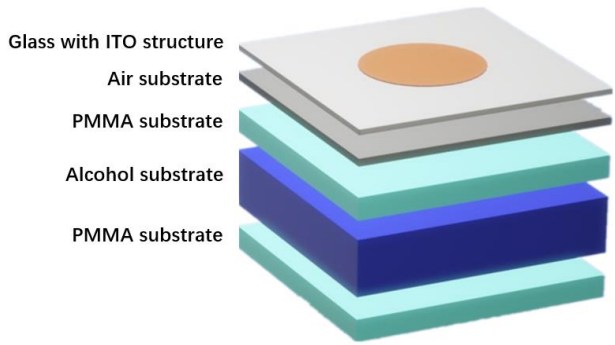

**Figure 1**. Basic composition schematic of a single unit cell.

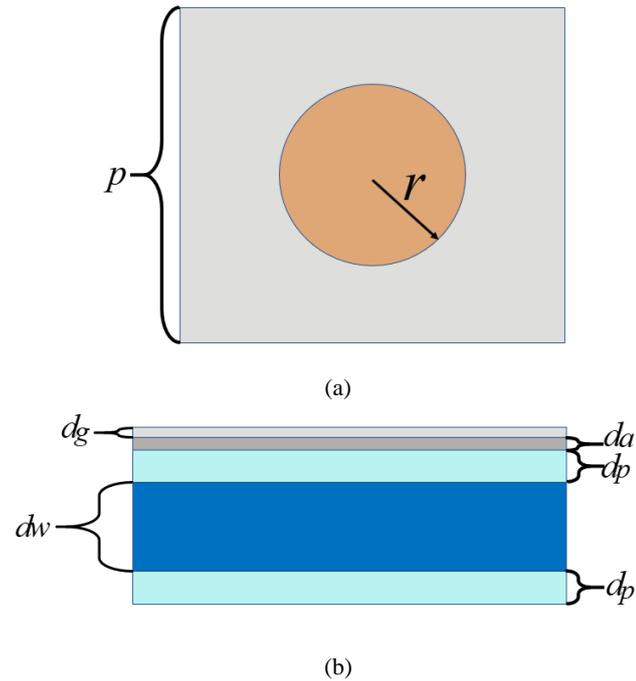

(a)

(b)

**Figure 2**. structure parameters of the unit cell.
(a) Front view, (b) Side view.

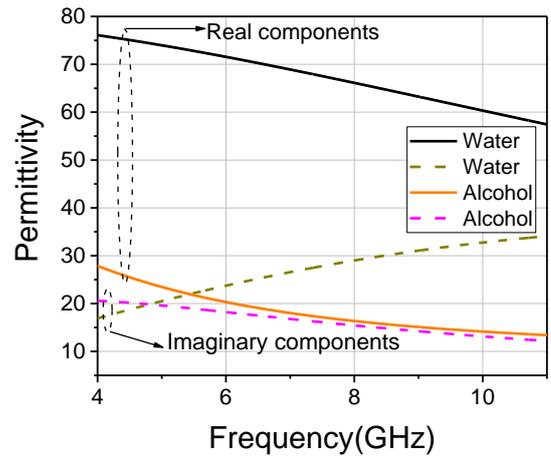

**Figure 3**. A comparison of the dielectric permittivity between water and alcohol at 20 ℃.

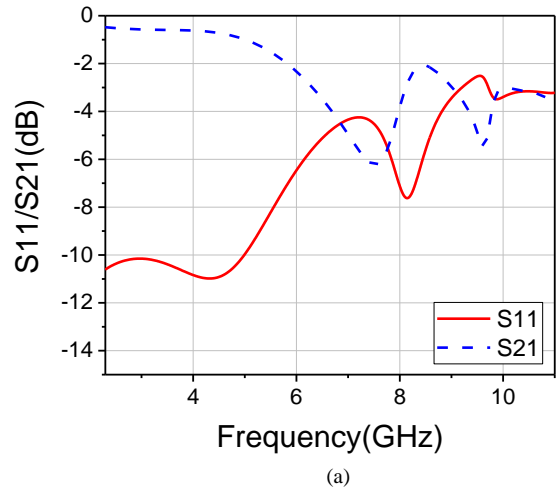

(a)

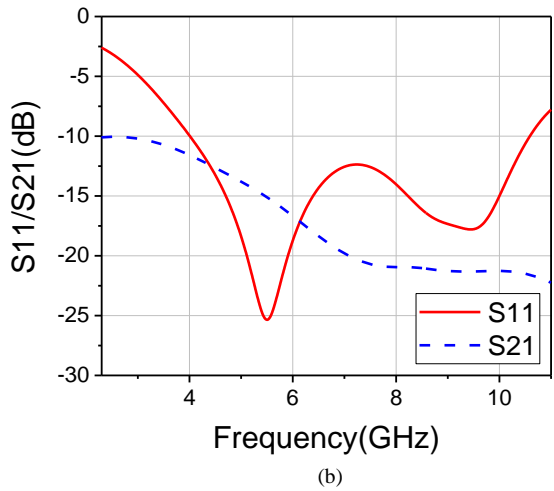
(b)

**Figure 4**. Reconfigurable performance of the window.
(a) Without alcohol, (b) with alcohol.

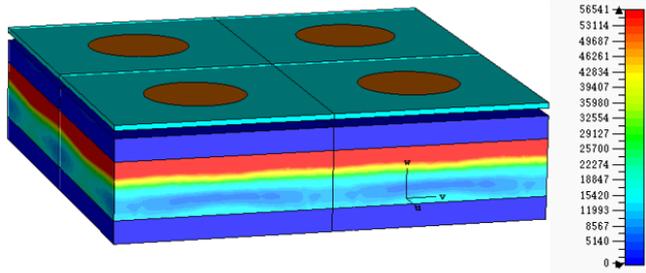

**Figure 5**. Power-loss density distribution of the window at 5.6GHz of absorbing state.

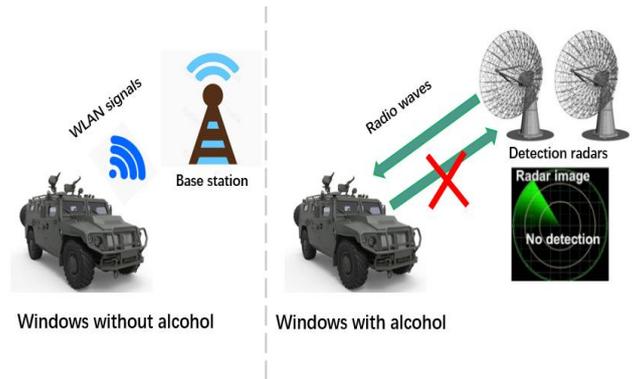

**Figure 6**. Illustration of an application scenario of the proposed window in a vehicle-mounted communication system.

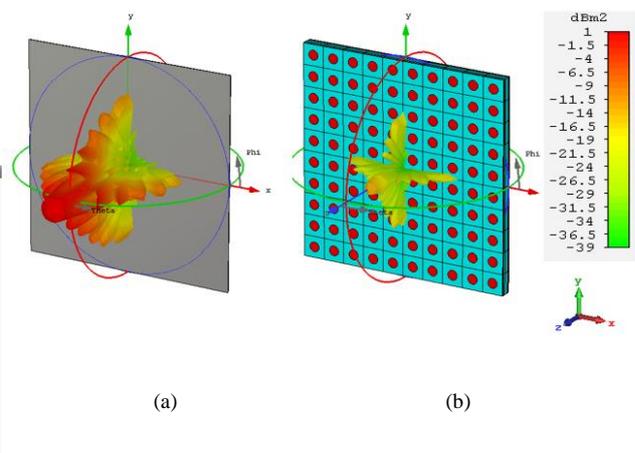

(a)                                    (b)

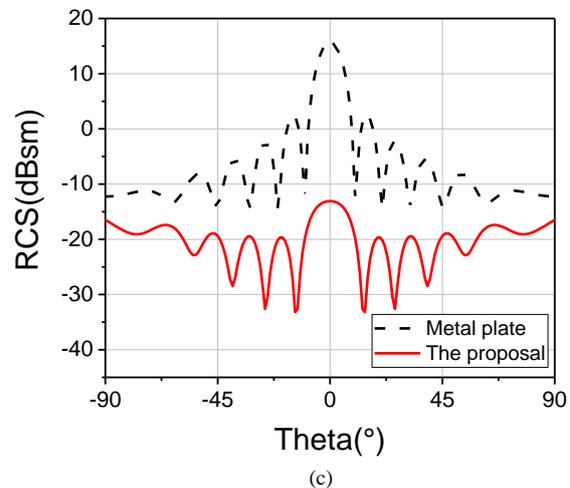

(c)

Figure 7 Simulated scattering patterns of PEC and the window at 5.6GHz of absorbing state.
(a) The pattern of PEC, (b) The pattern of the window, **(c)** far-field cut at $\varphi = 90°$.

where the values for the parameters $\varepsilon_\infty$, $\alpha_0$ and $\beta_0$ are obtained and shown in Table 1. It is necessary to mention that for different alcohol concentrations, the freezing point of the liquid is different. The formulae only work when the liquid is not frozen [35]. The permittivity of water [27] and 75% alcohol at 20°C at radiofrequency can both be described by the Debye formula. The real and imaginary parts of the permittivities of the water and alcohol mentioned above at 20 °C are shown in Figure 3. On the one hand, the absorption at radio frequency can be obtained because of the high imaginary part of the alcohol. On the other hand, according to the picture in Figure 3, the real part of the alcohol's permittivity is less than that of the water, making that geometrical parameter $d_g$ can be designed larger and fabricated easier. Therefore, we chose alcohol as the liquid layer.

As Figure 4(a) shown, the window can achieve a transmission passband from 2.3 GHz to 5 GHz with low insertion loss, especially at 2.45 GHz and 5 GHz with the insertion loss of the passband reaches 0.51 $dB$ and 0.99 $dB$, by alcohol drainage. In addition, the electromagnetic characteristic of the window without alcohol is studied, shown in Figure 4(b). It can half reflect electromagnetic waves at 2.45 GHz and absorb them from 4 GHz to 10.5 GHz with a strong absorptivity by alcohol injection. In this state, alcohol plays a significant part in absorbing electromagnetic waves. As depicted in Figure 5, most of the power from electromagnetic waves accepted by the window is absorbed by alcohol, which indicates that the alcohol is the main contributor to the broadband absorption.

To better understand the practical application of this proposal, an example of the window employed in a vehicle-mounted communication system is shown in Figure 6. In a realistic scenario, the electromagnetic environments are complex, making that the windows of vehicles should have multiple functions, such as a reconfigurable characteristic. In some domains, just depicted on the left side of Figure 6, a vehicle just as a stealth armored car needs to communicate with the base station for transmitting vital information, which makes that the window without alcohol can be employed in this vehicle to meet the need of good WLAN communication. Furthermore, revealed in the right side of Fig. 6, the stealth armored car with its windows filled with alcohol, changing from a transmission state to an absorption state, can absorb radio waves from 4 GHz to 10.5 GHz in some areas with detection radars installed, which have it avoid the risk of being detected. Therefore, in practical electromagnetic environments, the vehicles installed this liquid reconfigurable stealth window constructed by a metamaterial absorber can switch between the transmission state and the absorption state for satisfying different needs.

As for the absorption state, a detailed simulation was made to reveal good performance of absorbing electromagnetic waves. Figure 7(a) and Figure 7(b) show 3D scattering patterns of the window and the metal plate of the same size. Moreover, as depicted in Figure 7(c), the far-field patterns cut at φ = 90° between the metal plate and the window are made a comparison. It can be seen that the RCS of the proposed window has greatly reduced. To further demonstrate the characteristic of RCS reduction, a mono-static RCS simulation comparison among the metal plate, window without liquid and window with liquid was made in Figure 8 as supplementary.

## 3.Analysis of geometrical parameters and stability

To better reveal the absorption mechanism, the correlative parameters of the structure are studied. As shown in Figure 9(a)and Figure 9(b), different values of surface resistance from 20 $\Omega/Sq$ to 30 $\Omega/Sq$ of the top layer are simulated. It can be seen that a transmission passband from 2.3 GHz to 5 GHz with low insertion loss by alcohol drainage and an absorption performance from 4 GHz to 10.5 GHz by alcohol injection can be maintained. It is important to ensure the stability of the characteristic performance of our proposal as the values of surface resistance of the ITO

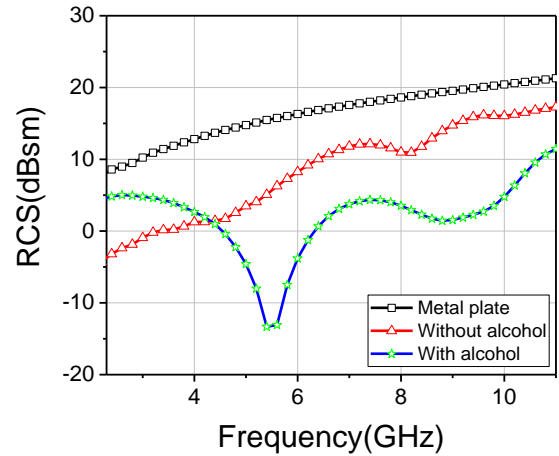

**Figure 8**. Simulated mono-static RCS of the window without and with alcohol and reference metal plate (θ = Φ = 0)

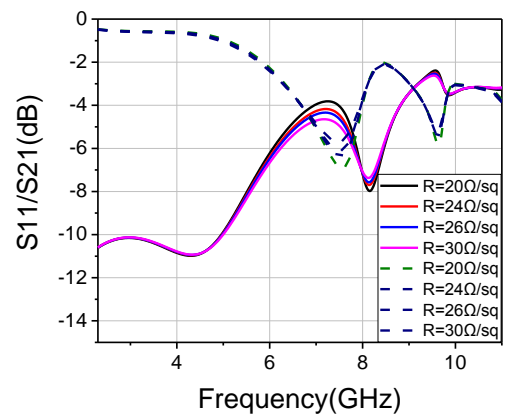

(a)

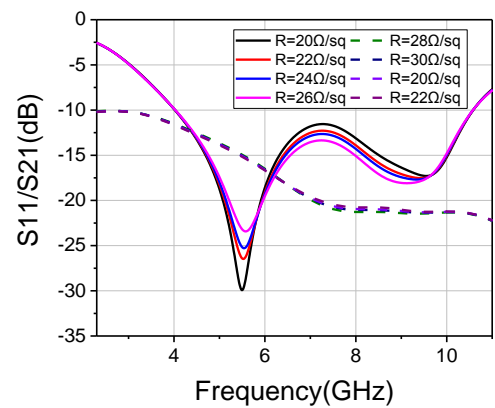

(b)

**Figure 9**. $S$ -parameters of the proposal with varied surface resistance of the ITO. (a)Without alcohol, (b) With alcohol.

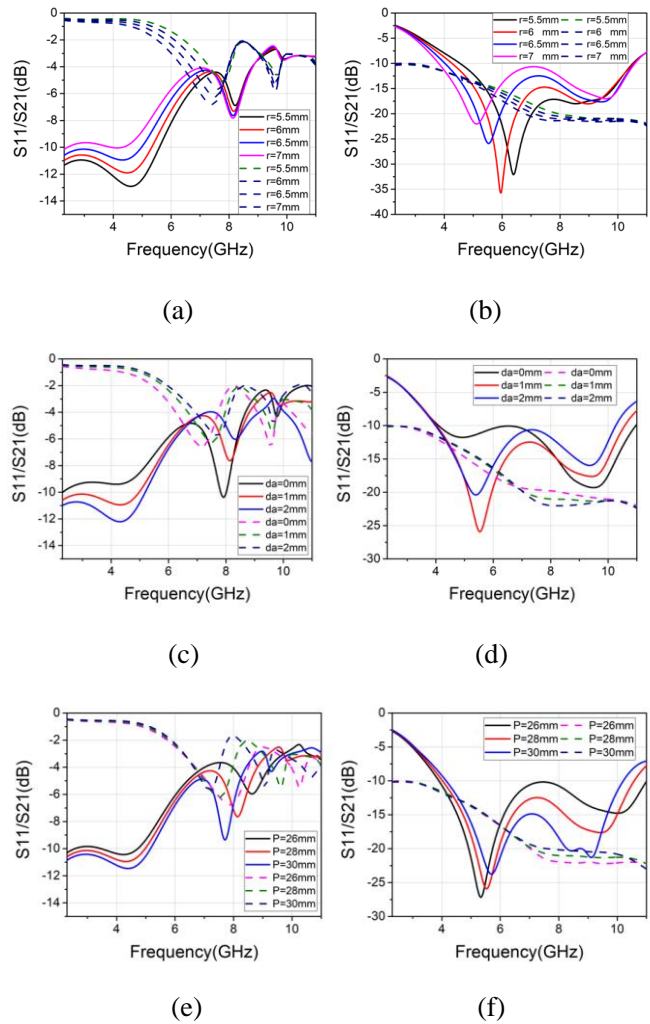

(a) (b)

(c) (d)

(e) (f)

**Figure 10**. Simulated sensitive parameters of the window. (a) (c) (e) Without alcohol, (b) (d) (f) With alcohol.

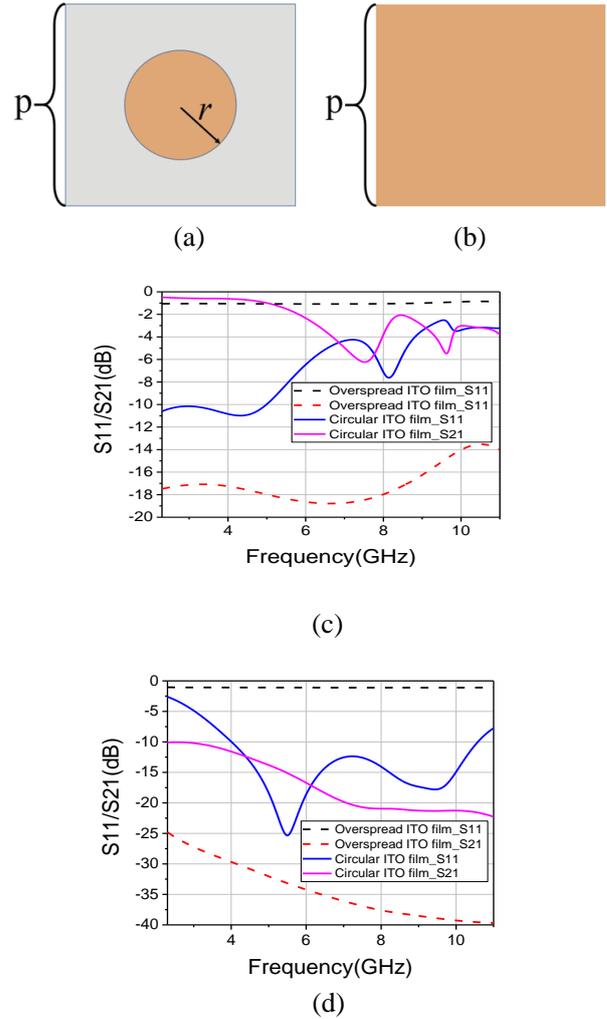

(a) (b)

(c)

(d)

**Figure 11**. The top layer with circular ITO film or no specific ITO film and the $S$-parameters.

(a) Circular ITO film, (b) overspread ITO film, The comparison of the S-parameters of the window (c) without and (d) with alcohol between circular ITO film and overspread ITO film.

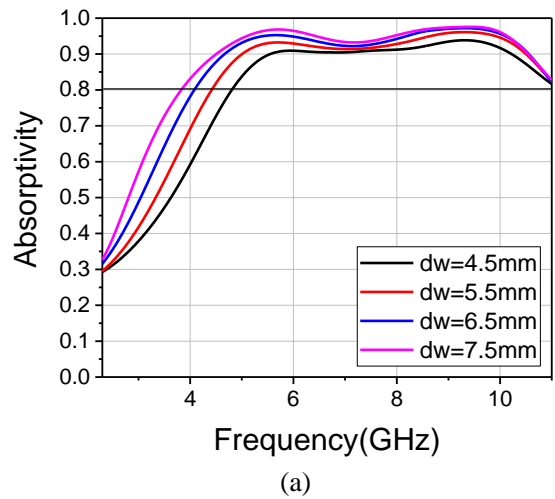

(a)

film change within a certain range because additional processing technology is not able to get the precise surface resistance, which allows a certain error range of manufacture.

From Figure 10, a liquid reconfigurable stealth window can be obtained with optimized parameters. A poor transmission passband will appear by alcohol drainage when these geometrical parameters deviate from optimized values. Furthermore, the geometrical parameters we designed can ensure the broadest absorption bandwidth with a strong absorptivity. In addition, as shown in Figure 11(a) and (b), we can see that there is no transmission passband or absorption band in the numerical simulation results while the surface of the glass is overspread with ITO film instead of the circular ITO structure, depicting that it is necessary to design a circular ITO structure film printed on the anti-reflection glass.

Journal XX (XXXX) XXXXXXAuthor *et al*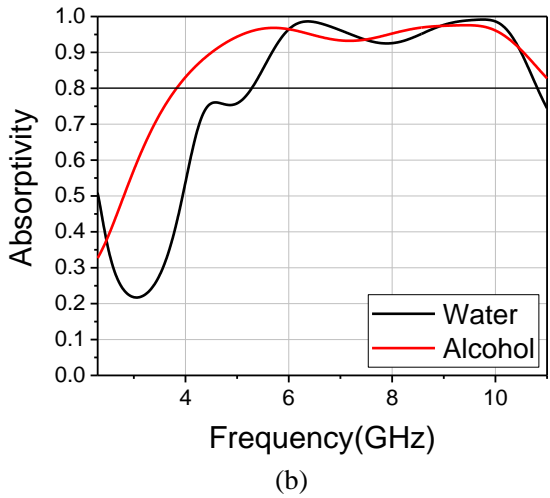

(b)

**Figure 12**. Absorptivity results of the window under (a)different thicknesses of the alcohol layer, (b) different liquid types.

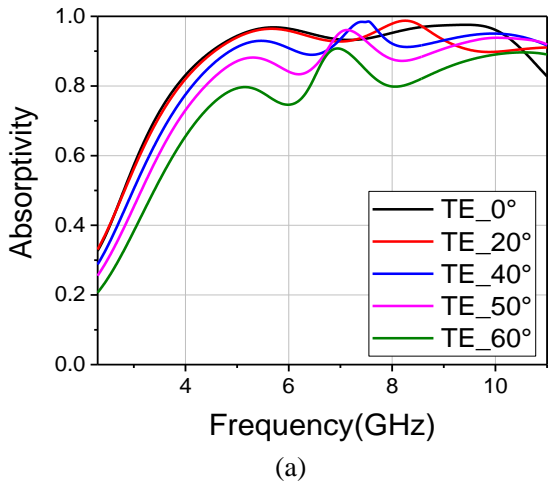

(a)

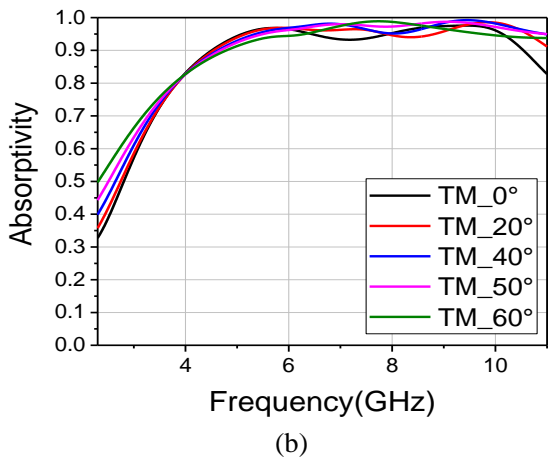

(b)

**Figure 13.** Absorptivity results under the oblique incidence of (**a**) TE, (**b**) TM polarization.

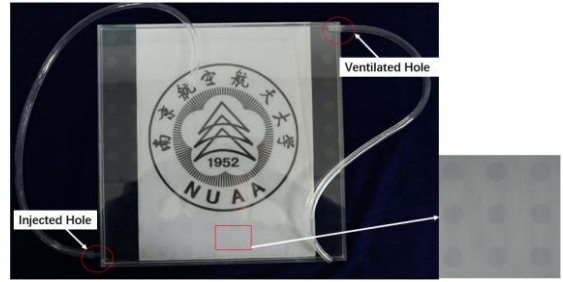

**Figure 14.** Propotype of the window.

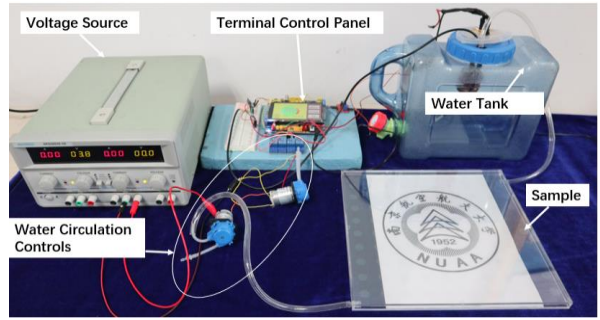

**Figure 15.** Liquid reconfigurable control system.

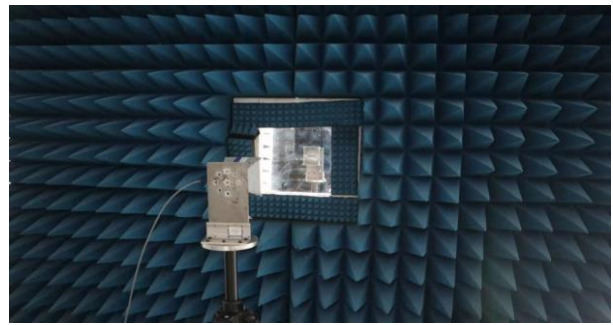

**Figure 16.** $S$-parameters measurement setup.

The proposal can be applied more flexibly in a realistic situation because its absorption band can be tunable. As shown in Figure 12(a), the numerical simulation results can be get by changing the thickness of the alcohol layer, in which only the TE wave with normal incidence is considered and the absorptivity can be worked as $A = 1 - |S_{11}|^2 - |S_{21}|^2$. It shows that the absorption bandwidth decreases with the decrease of the thickness of the alcohol layer, exhibiting that the absorption band is tunable. Moreover, as depicted in Figure 12(b), by changing the liquid type from alcohol to water with the fixed thickness of the liquid layer, the absorption band exhibits useful tunability distinctly. The tunability of the absorption band of the window makes it more practical and more flexible, having it more initiative in complex and changeable electromagnetic environments.

As shown in Fig. 13, the absorptivity change works as a function of the oblique incident angle. When the incident angle increases from 0° to 60°, the absorptivity is deteriorated gradually under TE polarization, resulting from

**Table 2.** Performances comparison with other transparent wideband absorbers.

| Reference | Absorption bandwidth, GHz (absorptivity) | Thickness (@ $f_L$) | Layers of ITO structure | Optical transmittance | Reconfigurable |
|---|---|---|---|---|---|
| [23] | 8.3-17.7(72.3%) | 0.08 $\lambda_L$ | 3 | 90%(Max.) | No |
| [24] | 8.3-17.4(70.8%) | 0.16 $\lambda_L$ | 2 | 77% | No |
| [25] | 3.5-16.6(130.3%) | 0.10 $\lambda_L$ | 3 | 60% | No |
| [29] | 7.5-15(66.7%) | 0.07 $\lambda_L$ | - | - | No |
| [30] | 6.4-23.7(115.0%) | 0.07 $\lambda_L$ | 2 | 71.1% | No |
| This work | 4.0-10.5(89.7%) | 0.20 $\lambda_L$ | 1 | 80.3% | Yes |

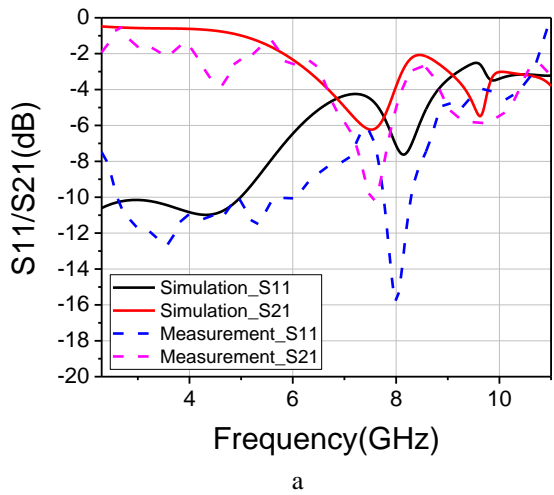

a

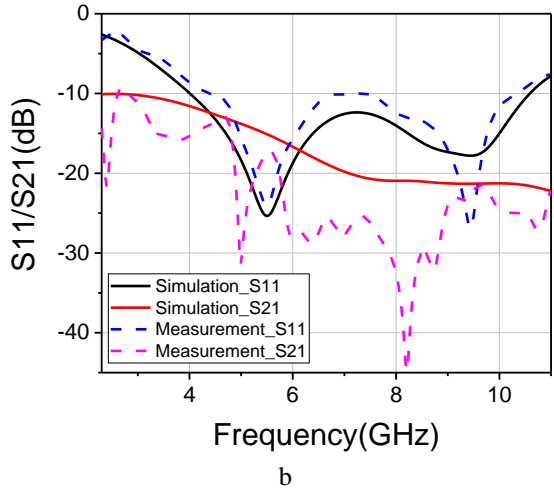

b

**Figure 17.** Experimental S-paraments results of the prototype. (a) withou alcohol, (b) with alcohol.

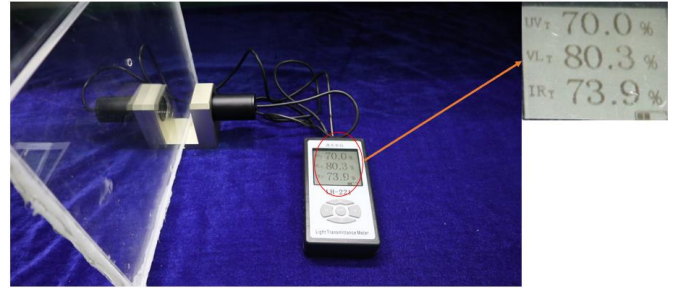

**Figure 18.** Transmittance test of the window.

the mismatched impedence with increase of the incident angle. But under TM polarization, the absorption performance maintains well with the increase of the incident angle from 0° to 60°.

## 4. Experiment results

A prototype of the proposed window is fabricated to validate the functions, as shown in Fig. 14. The liquid reconfigurable window is composed of 10×10 unit cells with an overall size of 286 *mm* × 286 *mm* × 15.2 *mm*. As for the top layer, magnetron sputtering is applied to print conductive ITO film on the glass, making the optical transparently anti-reflection glass with ITO film on it. Owing to the restriction of processing conditions and our comprehensive consideration, the glass substrate can only be fabricated with a thickness of 0.7 *mm*. The liquid container is constructed by PMMA to form a sealed space for filling with the alcohol layer. And some transparent small cylinders with 6 mm diameter and 1 mm thickness, made of PMMA, are used between the glass layer and the liquid container as supports to form an air layer. For filling and draining alcohol, two hoses are connected to two holes of the liquid container, respectively.

As shown in Figure 15, a liquid reconfigurable circulation system is set up to realize the reconfigurable function of the window conveniently. The control system consists of two parts: Part 1 is a terminal control panel composed of a single chip microcomputer and a touch screen;

Part 2 includes two independent water circulation controls comprising a voltage source, two water pumps, two relays, a water tank and two water pipes. We just use one of the water pumps to realize the injection or drainage of alcohol.

The transmission/absorption experiment was performed in a microwave anechoic chamber by the free-space method and the $S_{11}/S_{21}$ parameter was measured by an Agilent 5245A vector network analyzer (VNA). As shown in Figure 16, the measurement environment was set up and the prototype surrounded by absorbing materials was set in the middle of the measurement platform. Two horn antennas were connected to the VNA and fixed on the arms of the trestle, of which one is as a transmitter and the other is as a receiver. Two horn antennas were used in the measurement to cover $2-18\ GHz$ wavebands.

As shown in Figure 17, the measurement results of the prototype with and without alcohol at 20℃ are exhibited, compared with the simulated results. It can be seen that the comparison of the measured and simulated S-parameters of the window are in agreement basically. A slight deviation is obtained in the experimental results, which are probably caused by the fabrication error as well as the environmental noise during testing. Transparent small cylinders, as supports between the glass layer and the liquid container to form an air layer, maybe one of the factors causing the experimental error.

The optical transmittance is measured by the transmittance tester(LH-221), as shown in Figure 18 and the deviation of the instrument is <1%. The measured transmittance of visible light is 80.3% through random point measurement, which indicates that the proposal can be optical transparency even filled with alcohol.

To validate the performances of the window, some optically transparent absorbers are list to make a comparison in Table 2. As can be seen by comparing, the proposed window possesses high optical transparency, reconfigurable characteristic and simpler structure fabrication.

## 5.Conclusion

In this paper, a liquid reconfigurable stealth window constructed by metamaterial absorber is proposed. It is composed of anti-reflection glass with conductive ITO film, PMMA and alcohol. The proposal can obtain a transmission passband from 2.3 GHz to 5 GHz with low insertion loss by alcohol drainage and an absorption band from 4.5 GHz to 10.5 GHz with a strong absorptivity over 90% by alcohol injection, which indicates that the proposed transparent design have the reconfigurable characteristic. The numerical simulation and experimental results are in good agreement, demonstrating the reliability of our design.

## Acknowledgments

This work was supported in part by National Natural Science Foundation of China under Grant 62071227, in part by National Science Foundation of Jiangsu Province of China under Grant BK20201289, in part by Open Research Program in China's State Key Laboratory of Millimeter Waves under Grant K202027, in part by the Postgraduate Research & Practice Innovation Program of Jiangsu Province under Grant SJCX20_0070 and in part by the Fundamental Research Funds for the Central Universities under Grant kfjj20200403.